\documentclass[journal,12pt,onecolumn,draftclsnofoot,]{IEEEtran}
\usepackage[cmex10]{amsmath,mathtools} 
\makeatletter
\newcommand{\svast}{\bBigg@{3}}
\newcommand{\vast}{\bBigg@{4}}
\newcommand{\Vast}{\bBigg@{5}}

\makeatother

\usepackage{amsthm, amssymb, amsfonts} 
\usepackage{bm} 
\usepackage[ruled]{algorithm2e} 
\usepackage{cite} 
\usepackage[pdftex]{graphicx} 
\usepackage{epstopdf} 
\usepackage{color}
\usepackage{url}
\usepackage[keeplastbox]{flushend}
\usepackage{enumitem} 
\usepackage[caption=false]{subfig} 

\usepackage{hyperref} 
\hypersetup{colorlinks=true, linkcolor=black, citecolor=black} 

\usepackage{cleveref} 
\crefname{equation}{eq.}{eqs.}
\crefname{figure}{fig.}{figs.}

\newcommand{\RNum}[1]{\MakeUppercase{\romannumeral#1}} 
\begin{document}

\title{Secrecy Outage Analysis of  Two-Hop Decode-and-Forward Mixed RF/UWOC Systems}

\author{
    Yi~Lou,~\IEEEmembership{Member,~IEEE,}
    Ruofan~Sun,
    Julian~Cheng,~\IEEEmembership{Senior~Member,~IEEE,}
    Donghu~Nie,
    and~Gang~Qiao,~\IEEEmembership{Member,~IEEE}
}

\maketitle

\begin{abstract}
    We analyze the secrecy performance of a two-hop mixed radio frequency (RF)/underwater wireless optical communication (UWOC) system using a decode-and-forward (DF) relay. All RF and UWOC links are modeled by the $\alpha-\mu$ and exponential-generalized Gamma distributions, respectively. We first derive the expressions of the secrecy outage probability (SOP) in exact closed-form, which are subsequently used to derive asymptotic expressions at high SNR that only includes simple functions for further insight. Moreover, based on the asymptotic expression, we can determine the optimal transmit power for a wide variety of RF and UWOC channel conditions. All analyses are validated using Monte Carlo simulation.
\end{abstract}


\IEEEpeerreviewmaketitle

\section{Introduction}
\IEEEPARstart{U}{nderwater} wireless optical communication technology (UWOC) is emerging as an effective solution to the explosive growth of underwater applications \cite{zengSurveyUnderwaterOptical2017}. By using blue and green light, which have minimal attenuation when transmitting underwater, underwater optical communication can achieve ultra-high data rates over certain distances.

A two-hop communication system using a single relay is an effective means to extend the communication distance and improve the performance of the communication system. According to the forwarding mechanism used in the relay, the types of relays can be divided into amplify-and-forward (AF) relay and decode-and-forward (DF) relay.

To enable ultra-high-speed communication across the sea surface between underwater and airborne nodes, RF and UWOC technologies are often used in combination to form so-called hybrid RF/UWOC systems for ultra-high-speed transmissions across the sea surface between underwater and airborne nodes. With an ocean buoy or surface vessel acting as a relay, an  RF/UWOC communications system can be conveniently implemented in a two-hop configuration.

Physical layer security has been studied extensively in RF/FSO hybrid communication systems \cite{leiPerformanceAnalysisPhysical2016,leiSecrecyOutageAnalysis2018,leiSecrecyOutageAnalysis2018a}. Recently, the security issue of hybrid RF/UWOC communication systems has become a hot topic of research.  The secrecy performance of two-hop mixed RF/UWOC systems using AF \cite{illiDualHopMixedRFUOW2018} and DF \cite{illiPhysicalLayerSecurity2020} relays are both studied, where the RF channel is modeled using a Nakagami distribution and the UWOC channel is modeled using a mixture exponential-Gamma distribution.

However, some authors have recently proposed a more accurate distribution, i.e., exponential-generalized Gamma (EGG), for modeling UWOC channels through laboratory experiments, which can take into account not only the temperature gradient but also the effect of bubbles on turbulence, for both freshwater and salty water \cite{zediniUnifiedStatisticalChannel2019}. Further,  Nakagami distribution is not general enough to model more realistic physical fading scenarios. The $\alpha-\mu$ distribution is a more flexible channel model that can model more realistic physical scenarios using two distribution parameters, $\alpha$ and $\mu$, to describe the non-linearity of the propagation medium and the number of clusters of multipath waves, respectively \cite{yacoubAmDistributionPhysical2007}. Moreover, The $\alpha-\mu$ distribution can be easily extended to Rayleigh, Nakagami-$m$, Weibull, one-sided Gaussian, etc., by setting the parameters $\alpha$ and $\mu$ to specified values.

This paper is the first to analyze the secrecy performance of a mixed RF/UWOC system using a DF relay, where the UWOC link is modeled using the novel EGG distribution and all the RF links are modeled using the versatile $\alpha$-$\mu$ distributions. We derive the exact closed-expression of the secrecy outage probability (SOP) in terms of bivariate $H$-functions. Moreover, for further enlightenment and determine the optimal transmitting power, we derive the asymptotic SOP expressions at high SNR that includes only simple functions.

The rest of this letter is organized as follows. In Section \RNum{2}, the channel and system models are presented. In Section \RNum{3}, the end-to-end performance metrics are studied. Numerical results are discussed in Section \RNum{4}, followed by the conclusion in Section \RNum{5}.

\section{System and Channel Models}
We consider a mixed RF/UWOC system in which the source node (S) in the air transmits its private data to the legitimate destination node (D) located underwater via a trusted DF relay node (R), which can be a buoy or a surface ship. RF channel from source to relay and underwater optical channel from the relay to destination node are assumed to follow $\alpha-\mu$ and EGG distributions, respectively. One unauthorized receiver (E) attempts to eavesdrop on RF signals received by the relay node during transmission.
\subsection{RF channel model}
All the RF links are assumed to be block fading and modeled by the $\alpha$-$\mu$ distribution. The probability density function (PDF) for the instantaneous SNR of SR link (denoted by $\gamma_1$) and SE link (denoted by $\gamma_e$) are given as $\gamma_{k}=\frac{P_{k}}{\sigma_{k}^{2}}=\bar{\gamma}_{k} |h_k|^2$, where $k\in\{1,e\}$, $|h_k|^2$ denotes the instantaneous channel power gain, $P_k$ denotes the transmission power, and $\sigma_{k}^{2}$ denotes the noise power. $\gamma_{k}$ can be expressed as \cite{kongPhysicalLayerSecurity2019}
\begin{IEEEeqnarray*}{rcl}
    f_{\gamma _k}\left(\gamma _k\right)&=&\frac{\alpha }{\Gamma (\mu )}\frac{\mu ^{\mu }}{\left(\bar{\gamma }_k\right)^{\alpha \mu}}\gamma _k^{\alpha \mu-1}\thinspace\text{exp}\left(-\mu \left(\frac{\gamma _k}{\bar{\gamma }_k}\right)^{\alpha}\right) \IEEEyesnumber \label{pdfam}
\end{IEEEeqnarray*}
where $k\in\{1,e\}$, $\mu \geq 0, \alpha \geq 0, \gamma_k \geq 0$, and $\Gamma(\cdot)$ is the gamma function. The distribution parameters $\alpha$ and $\mu$ dictate the non-linearity and multipath propagation characteristics of the fading model.

Using \eqref{pdfam}, the complementary cumulative distribution function (CCDF) of $\gamma_1$, i.e., $\overline{F}_{\gamma_1}(\gamma)$ can be derived as follows
\begin{IEEEeqnarray*}{rcl}
    \bar{F}_{\gamma _1}\left(\gamma _1\right)&=&\int _{\gamma _1}^{\infty }f_{\gamma _1}\left(\gamma _1\right)d\gamma _1 \\
    &\overset{(a)}{=}&\int _{\gamma _1}^{\infty }\kappa  H_{0,1}^{1,0}\!\!\left[\gamma _1 \Lambda \middle|\!\!\!\begin{array}{c}   \\ \left(-\frac{1}{\alpha }+\mu ,\frac{1}{\alpha }\right) \\\end{array}\!\!\!\right]d\gamma _1 \\
    &\overset{(b)}{=}&-\frac{i \kappa  }{2 \pi }\int _{\mathcal{L}}^s\Lambda ^{-s} \Gamma \left(\frac{s}{\alpha }+\mu -\frac{1}{\alpha }\right) \int _{\gamma _1}^{\infty}\gamma _1^{-s}d\gamma _1ds \\
    &\overset{(c)}{=}&\gamma  \kappa  H_{1,2}^{2,0}\!\!\left[\gamma  \Lambda \middle|\!\!\!\begin{array}{c} (0,1) \\ (-1,1),(-\frac{1}{\alpha }+\mu ,\frac{1}{\alpha })  \\\end{array}\!\!\!\right]\IEEEyesnumber\label{ccdf1}
\end{IEEEeqnarray*}
where $\kappa=\frac{\beta}{\Gamma(\mu) \bar{\gamma}_{k}}$, $ \Lambda=\frac{\beta}{\bar{\gamma}_{k}}$, $\beta=\frac{\Gamma\left(\mu+\frac{1}{\alpha}\right)}{\Gamma(\mu)}$, and  $H_{\cdot ,\cdot }^{\cdot ,\cdot }[\cdot|\cdot]$ is the $H$-Function \cite[Eq. (1.2)]{mathaiHFunctionTheoryApplications2010}. Step $(a)$ is derived by using \cite[Eq. (1.125)]{mathaiHFunctionTheoryApplications2010}. Step $(b)$ is obtained by using \cite[Eq. (1.1.2)]{kilbasHtransformsTheoryApplications2004} and rearranging the integral variables.  Step $(c)$ is obtained by solving the integral with respect to x, and using \cite[Eq. (1.1.1) ]{kilbasHtransformsTheoryApplications2004} and \cite[Eq. (2.1.5) ]{kilbasHtransformsTheoryApplications2004}.

\subsection{UWOC channel model}
To consider the combined effects of air bubbles and temperature gradients of the UWOC channel on the received optical irradiance in both pure water and salty water, we modeled the UWOC fading using the EGG distribution. The instantaneous SNR D of a IM/DD-based system with OOK modulation is defined as $\gamma_2=\left(\eta I\right)^{2} / N_{0}$, where $\eta$ is the effective photoelectric conversion ratio, and $N_{0}$ denotes the power of noise\cite{zediniUnifiedStatisticalChannel2019}. the PDF of $I$ can be expressed as \cite{zediniUnifiedStatisticalChannel2019}
\begin{IEEEeqnarray*}{rcl}
    f_{I}\left(I\right) =\frac{\omega}{\lambda} \exp \left(-\frac{I}{\lambda}\right) +\left(1-\omega\right) \frac{c I^{a c-1}}{b^{a c} \Gamma\left(a\right)} \exp \left(-\frac{I^{c}}{b^{c}}\right)\thinspace\thinspace\thinspace\thinspace\IEEEyesnumber
\end{IEEEeqnarray*}
where $\omega$ is the mixture weight of the EGG distribution, $\lambda$ is the parameter linked to the exponential distribution, $a$, $b$ and $c$ are the parameters related to the exponential distribution.

The PDF of the instantaneous received SNR at D can be given as
\begin{IEEEeqnarray*}{rcl}
    f_{\gamma_2}(\gamma)&=&\frac{c(1-\omega)}{\gamma r \Gamma(a)} e^{-\left(\frac{\gamma}{b^{r} \mu_r}\right)^{\frac{c}{r}}}\left(\frac{\gamma}{b^{r} \mu_{r}}\right)^{\frac{a c}{r}}\\
    &+&\frac{\omega}{\gamma \lambda r}\left(\frac{\gamma}{\mu_{r}}\right)^{\frac{1}{r}} e^{-\frac{1}{\lambda}\left(\frac{\gamma}{\mu_{r}}\right)^{\frac{1}{r}}}.\IEEEyesnumber
\end{IEEEeqnarray*}

The CCDF of $\gamma_2$ is therefore obtained from \cite[Eq. (3.381.3)]{i.s.gradshteynTableIntegralsSeries2007}, and is given as
\begin{IEEEeqnarray*}{rcl}
    \bar{F}_{\gamma _2}(\gamma _2)&=&\int _{\gamma _2}^{\infty }f_{\gamma _2}\left(\gamma _2\right)d\gamma _2 \\
    &=&\omega\!  \thinspace\text{exp}\left(-\frac{1}{\lambda }\right)\left(\frac{\gamma }{\mu _r}\right)^{\frac{1}{r}}\!-\!\frac{(\omega -1)}{\Gamma (a)}\Gamma \!\left(\!a, \frac{\gamma^{c/r}}{b^c\mu _r^{c/r}}\!\right)\IEEEyesnumber\label{ccdf2}
\end{IEEEeqnarray*}
where $\gamma(\cdot,\cdot)$ is the upper incomplete Gamma function \cite[Eq. (8.350.2)]{i.s.gradshteynTableIntegralsSeries2007}.

\section{SOP}
An SOP defines the probability of failing to obtain a reliable and secure transmission. SOP is the most commonly used performance metric for evaluating the secrecy performance of communication systems in the presence of eavesdroppers\cite{research.WolframFunctionsSite2020}, and can be expressed as
\begin{IEEEeqnarray*}{rcl}
    P_{\text {out}}\left(R_{s}\right)=\operatorname{Pr}\left\{C_{s}\left(\gamma_{e q}, \gamma_{e}\right) \leq R_{s}\right\}.\IEEEyesnumber
\end{IEEEeqnarray*}

Referring to \cite{leiSecrecyPerformanceMixed2017}, the lower bound for the SOP is derived as
$$ P_{\text {out}}^L\left(R_{s}\right) \approx\int_{0}^{\infty} F_{\gamma_{e q}}(\Theta \gamma) f_{\gamma_{e}}(\gamma) d \gamma$$
where $\gamma_{e q}$ is the end-to-end instantaneous SNR of the mixed RF/UWOC system using the DF relay, and can be given as follows
\begin{IEEEeqnarray*}{rcl}
    \gamma_{eq}=\min \left(\gamma_{1}, \gamma_{2}\right).\IEEEyesnumber \label{snreqstep1}
\end{IEEEeqnarray*}

Using \eqref{snreqstep1}, the CDF of the SNR $\gamma_{eq}$ can be expressed as
\begin{IEEEeqnarray*}{rcl}
    F_{\gamma_{e q}}(\gamma) &=&\operatorname{Pr}\left[\min \left(\gamma_{1}, \gamma_{2}\right)<\gamma\right] \\
    &=&1-\left(1-F_{\gamma_{1}}(t)\right)\left(1-F_{\gamma_{2}}(\gamma)\right) \\
    &=&1-\bar{F}_{\gamma _2}(\gamma) \bar{F}_{\gamma _1}(\gamma). \IEEEyesnumber \label{eq1}
\end{IEEEeqnarray*}

After substituting \eqref{ccdf1} and \eqref{ccdf2} into \eqref{snreqstep1} and some simplification,  $\gamma_{eq}$ can be expressed in the following form
\begin{IEEEeqnarray*}{rcl}
    &&F_{\text{eq}}(\gamma )=1+\frac{\kappa  (\omega -1)\text{  }}{\Lambda  \Gamma (a)}\Gamma \left(a,b^{-c} \left(\frac{\gamma }{\mu _r}\right)^{c/r}\right)\\
    &&\times H_{1,2}^{2,0}\left[\gamma \Lambda \left|\begin{array}{c} (1,1) \\ (0,1),(\mu ,\frac{1}{\alpha }) \\\end{array}\right.\right]\!-\!\frac{\kappa  \omega }{\Lambda }\thinspace\text{exp}\left(-\frac{1}{\lambda }\left(\frac{\gamma }{\mu _r}\right)^{\frac{1}{r}}\right)\\
    &&\times H_{1,2}^{2,0}\left[\gamma  \Lambda\left|\begin{array}{c} (1,1) \\ (0,1),(\mu ,\frac{1}{\alpha }) \\\end{array}\right.\right]. \IEEEyesnumber\label{cdfeq}
\end{IEEEeqnarray*}

Using \eqref{pdfam} and \eqref{cdfeq} and after some simplification, $SOP_L$ can be expressed as follows
\begin{IEEEeqnarray*}{rcl}
    P_{\text {out}}^L&=&1+K_1+K2\IEEEyesnumber\label{sopstep1}
\end{IEEEeqnarray*}
where
\begin{IEEEeqnarray*}{rcl}
    &&K_1=\frac{\kappa  (1-\omega ) \alpha _e \kappa _e\text{  }}{\Lambda  \Gamma (a)}\Lambda _e^{\alpha _e \mu _e-1}\int _0^{\infty }\thinspace\text{exp}\left(-\left(\gamma \Lambda _e\right)^{\alpha _e}\right)\gamma ^{\alpha _e \mu _e-1}\\
    &&\times  \Gamma \left(a,b^{-c}\left(\frac{\gamma  \Theta }{\mu _r}\right)^{\frac{c}{r}}\right)H_{1,2}^{2,0}\left[\gamma  \Theta  \Lambda \left|\begin{array}{c} (1,1) \\ (0,1),(\mu ,\frac{1}{\alpha }) \\\end{array}\right.\right]d\gamma\IEEEyesnumber \label{k1step1}
\end{IEEEeqnarray*}
and
\begin{IEEEeqnarray*}{rcl}
    K_2&=&-\frac{\kappa  \omega  \alpha _e \kappa _e }{\Lambda }\Lambda _e^{\alpha _e \mu _e-1}\!\int _0^{\infty }\!\!\thinspace\text{exp}\left(\!-\left(\gamma  \Lambda_e\right)^{\alpha _e}\!-\!\frac{1}{\lambda }\left(\frac{\gamma  \Theta }{\mu _r}\right)^{\frac{1}{r}}\!\right)\\
    &\times& \gamma ^{\alpha _e \mu _e-1} H_{1,2}^{2,0}\left[\gamma \Theta  \Lambda \left|\begin{array}{c} (1,1) \\ (0,1),(\mu ,\frac{1}{\alpha }) \\\end{array}\right.\right]d\gamma.\IEEEyesnumber\label{k2step1}
\end{IEEEeqnarray*}

To obtain a closed expression for SOP, we use $H$ function to express the exponential functions in \eqref{k1step1} and \eqref{k2step1} as \cite[Eq. (07.34.03.0046.01)]{mmafunction}
\begin{IEEEeqnarray*}{rcl}
    \thinspace\text{exp}\left(-\left(\gamma  \Lambda _e\right)^{\alpha _e}\right)=\frac{1}{\alpha _e}H_{0,1}^{1,0}\!\!\left[\gamma  \Lambda _e\middle|\!\!\!\begin{array}{c}   \\ \left(0,\frac{1}{\alpha _e}\right) \\\end{array}\!\!\!\right]\IEEEyesnumber\label{j1}
\end{IEEEeqnarray*}
and
\begin{IEEEeqnarray*}{rcl}
    \thinspace\text{exp}\left(-\frac{1}{\lambda }\left(\frac{\gamma  \Theta }{\mu _r}\right)^{\frac{1}{r}}\right)=r H_{0,1}^{1,0}\!\!\left[\frac{\gamma  \Theta  \lambda ^{-r}}{\mu_r}\middle|\!\!\!\begin{array}{c}   \\ (0,r) \\\end{array}\!\!\!\right].\IEEEyesnumber\label{j2}
\end{IEEEeqnarray*}

The Generalized gamma function in \eqref{k1step1} is also expressed in the form of $H$ function as \cite[Eq. (06.06.26.0005.01)]{mmafunction}
\begin{IEEEeqnarray*}{rcl}
    \Gamma \left(\!\!a,\!b^{-c}\!\left(\frac{\gamma  \Theta }{\mu _r}\right)^{\frac{c}{r}}\!\right)\!\!=\!\!H_{1,2}^{2,0}\!\!\left[b^{-c}\left(\frac{\gamma  \Theta }{\mu_r}\right)^{\frac{c}{r}}\middle|\!\!\!\begin{array}{c} (1,1) \\ (0,1),(a,1) \\\end{array}\!\!\!\right].\IEEEyesnumber\label{j3}
\end{IEEEeqnarray*}

\begin{figure*}[!bht]
    \begin{IEEEeqnarray*}{rcl}
        K_1=-\frac{\kappa  r \omega  \kappa _e }{\Lambda  \Lambda _e}\left(\frac{\Lambda _e \lambda ^r \mu _r}{\Theta }\right)^{\alpha _e \mu _e}H_{1,0:0,1;1,2}^{0,1:1,0;2,0}\!\!\left[\begin{array}{c} \frac{\lambda ^r \Lambda _e \mu _r}{\Theta } \\ \lambda ^r \Lambda  \mu _r \\\end{array}\middle|\!\!\!\begin{array}{ccccc} \left(1-r \alpha _e \mu _e,r,r\right) & : &                                    & ; & (1,1)                          \\
                                                              & : & \left(0,\frac{1}{\alpha _e}\right) & ; & (0,1),(\mu ,\frac{1}{\alpha }) \\\end{array}\!\!\!\right]\IEEEyesnumber\label{k1exact}
    \end{IEEEeqnarray*}
    \hrule
    \begin{IEEEeqnarray*}{rcl}
        K_2=-\frac{\kappa  (1-\omega ) \kappa _e }{\Lambda  \Gamma (a) \Lambda _e}H_{1,0:1,2;1,2}^{0,1:2,0;2,0}\!\!\left[\begin{array}{c} b^{-c}\left(\frac{\Theta }{\Lambda _e \mu _r}\right)^{\frac{c}{r}} \\ \frac{\Theta  \Lambda }{\Lambda _e} \\\end{array}\middle|\!\!\!\begin{array}{ccccc} \left(1-\mu _e,\frac{c}{r \alpha _e},\frac{1}{\alpha _e}\right) & : & (1,1)       & ; & (1,1)                          \\
                                                                                        & : & (0,1),(a,1) & ; & (0,1),(\mu ,\frac{1}{\alpha }) \\\end{array}\!\!\!\right]\IEEEyesnumber\label{k2exact}
    \end{IEEEeqnarray*}
    \hrule
\end{figure*}

Then, using the mellin transform of the product of  three $H$-functions \cite[Eq. (2.3)]{mittalIntegralInvolvingGeneralized1972}, we can obtain the exact closed-form of $K_1$ and $K_2$ as in \eqref{k1exact} and \eqref{k2exact}, in terms of bivariate $H$-function, respectively. Finally, a closed-form expression for SOP can be readily deduced by substituting \eqref{k1exact} and \eqref{k2exact} into \eqref{sopstep1}. Also note that, bivariate $H$-function has already been implemented in MATLAB \cite{cherguiRicianKfactorbasedAnalysis2018}, Python \cite{alhennawiClosedFormExactAsymptotic2016}, and Mathematica \cite{almeidagarciaCACFARDetectionPerformance2019}, and can be easily evaluated.

In order to demonstrate the usefulness of the exact closed-form expression of  SOP and gain more insight into the effect of model parameters of the EGG and $\alpha-\mu$ channels on the secrecy performance, we next derive the asymptotic expression for SOP at high SNRs. We consider two scenarios,  i.e., $\gamma_1\to \infty$ and $\gamma_e\to \infty$.

We first consider the scenario $\gamma_1\to \infty$. Based on the definition of the bivariate $H$-function, we can rewrite $K_1$ as
\begin{IEEEeqnarray*}{rcl}
    &&K_1=\frac{\kappa  (1-\omega ) \kappa _e}{4 \pi ^2 \Lambda  \Gamma (a) \Lambda _e}\int _{\mathcal{L}}^t\int _{\mathcal{L}}^s\frac{\Gamma (a-s) \Gamma(-s) \Gamma (-t) \Gamma \left(\mu -\frac{t}{\alpha }\right)}{\Gamma (1-s) \Gamma (1-t)}\\
    &&\Gamma\! \left(\frac{c s}{r \alpha _e}\!+\!\mu _e\!+\!\frac{t}{\alpha_e}\!\right)\left(\frac{\Theta  \Lambda }{\Lambda _e}\right)^t \left(b^{-c} \Theta ^{c/r} \Lambda _e^{-\frac{c}{r}} \mu _r^{-\frac{c}{r}}\right)^sdsdt.\IEEEyesnumber \label{k1line}
\end{IEEEeqnarray*}

Observing that as $\gamma_1\to \infty$, $\frac{\Theta  \Lambda }{\Lambda _e}\to 0$, we therefore convert the line integral of $t$ in \eqref{k1line} into the form of the $H$-function. Then, $K_1$ can be rewritten as
\begin{IEEEeqnarray*}{rcl}
    K_1&\approx&2 i \pi \frac{\kappa  (1-\omega ) \kappa _e}{4 \pi ^2 \Lambda  \Gamma (a) \Lambda _e}\int _{\mathcal{L}}^s\frac{ \Gamma (a-s) \Gamma (-s)}{\Gamma(1-s)}b^{-\text{cs}}\left(\frac{\Theta }{\Lambda _e \mu _r}\right)^{\frac{c s}{r}}\\
    &\times&\underbrace{H_{2,2}^{2,1}\!\!\left[\frac{\Theta  \Lambda }{\Lambda _e}\middle|\!\!\!\begin{array}{c} \left(1-\frac{c s}{r \alpha _e}-\mu _e,\frac{1}{\alpha _e}\right),(1,1) \\ (0,1),(\mu ,\frac{1}{\alpha }) \\\end{array}\!\!\!\right]}_{\Xi _1}ds.\IEEEyesnumber
\end{IEEEeqnarray*}

Using \cite[Eq. (1.8.4)]{kilbasHtransformsTheoryApplications2004},  $\Xi _1$ can be asymptotically expanded as the sum of the residues of all poles to the left of the contour, and is given as
\begin{IEEEeqnarray*}{rcl}
    \Xi _1&\approx&\Gamma (\mu ) \Gamma \left(\frac{c s}{r \alpha _e}+\mu _e\right)-\frac{1}{\mu }\left(\frac{\Theta  \Lambda }{\Lambda _e}\right)^{\alpha  \mu}\\
    &\times& \Gamma \left(\frac{c s+r \alpha  \mu }{r \alpha _e}+\mu _e\right).\IEEEyesnumber
\end{IEEEeqnarray*}

After some simplification, we can express $K_1$ as $\gamma_1\to \infty$ into the following form
\begin{IEEEeqnarray*}{rcl}
    K_1&\approx&\left(\frac{\Theta  \Lambda }{\Lambda _e}\right)^{\alpha  \mu}H_{2,2}^{1,2}\!\!\left[b^c \left(\frac{\Lambda _e \mu _r}{\Theta }\right)^{\frac{c}{r}}\middle|\!\!\!\begin{array}{c} (1-a,1),(1,1) \\ \left(\frac{\alpha  \mu }{\alpha _e}+\mu _e,\frac{c}{r \alpha _e}\right),(0,1) \\\end{array}\!\!\!\right]\\
    &\times&\frac{(1-\omega )  }{\Gamma (a) \Gamma (\mu +1) \Gamma \left(\mu _e\right)}-\frac{(1-\omega )}{\Gamma (a) \Gamma \left(\mu _e\right)}\\
    &\times&H_{2,2}^{1,2}\!\!\left[b^c \left(\frac{\Lambda _e \mu _r}{\Theta }\right)^{\frac{c}{r}}\middle|\!\!\!\begin{array}{c} (1-a,1),(1,1) \\ \left(\mu _e,\frac{c}{r \alpha _e}\right),(0,1) \\\end{array}\!\!\!\right].\IEEEyesnumber \label{k1final1}
\end{IEEEeqnarray*}

Similarly, to obtain the asymptotic expression for $K_2$ as  $\gamma_1\to \infty$ , we first represent $K_2$ as the following form
\begin{IEEEeqnarray*}{rcl}
    K_2&=&\frac{\kappa  r \omega  \kappa _e}{4 \pi ^2 \Lambda }\Lambda _e^{\alpha _e \mu _e-1}\left(\frac{\Theta  \lambda ^{-r}}{\mu _r}\right)^{-\alpha_e \mu _e}\int _{\mathcal{L}}^t\int _{\mathcal{L}}^s\frac{\Gamma \left(\mu -\frac{t}{\alpha }\right)}{\Gamma (1-t)}\\
    &\times&\Gamma (-t) \Gamma \left(-\frac{s}{\alpha_e}\right)\Gamma \left(r s+r t+r \alpha _e \mu _e\right) \left(\lambda ^r \Lambda  \mu _r\right)^t \\
    &\times&\left(\frac{\lambda ^r \Lambda _e \mu _r}{\Theta}\right)^sdsdt.\IEEEyesnumber
\end{IEEEeqnarray*}

Noting that as $\gamma_1\to \infty$, $\lambda^{r}\Lambda  \mu _r\to \infty$, so we rewrite $K_2$ to the following form
\begin{IEEEeqnarray*}{rcl}
    K_2&\approx&\frac{i \kappa  r \omega  \kappa _e}{2 \pi  \Lambda  \Lambda _e}\left(\frac{\Lambda _e \lambda ^r \mu _r}{\Theta }\right)^{\alpha _e \mu _e}\int_{\mathcal{L}}^s\Gamma \left(-\frac{s}{\alpha _e}\right) \left(\frac{\lambda ^r \Lambda _e \mu _r}{\Theta }\right)^s\\
    &\times&\underbrace{ H_{2,2}^{2,1}\!\!\left[\lambda ^r \Lambda  \mu _r\middle|\!\!\!\begin{array}{c} \left(1-r \left(s+\alpha _e \mu _e\right),r\right),(1,1) \\ (0,1),(\mu ,\frac{1}{\alpha }) \\\end{array}\!\!\!\right]}_{\Xi _2}ds.\IEEEyesnumber\label{k2step2}
\end{IEEEeqnarray*}

Using \cite[Eq. (1.5.9)]{kilbasHtransformsTheoryApplications2004},  $\Xi _2$ can be asymptotically expanded as the sum of the residues of all poles to the right of the contour, and is given as
\begin{IEEEeqnarray*}{rcl}
    \Xi _2&\approx&\Gamma (\mu ) \Gamma \left(r \left(s+\alpha _e \mu _e\right)\right)-\frac{1}{\mu }\left(\Lambda  \lambda ^r \mu _r\right)^{\alpha  \mu }\\
    &\times& \Gamma\left(r \left(s+\alpha  \mu +\alpha _e \mu _e\right)\right).\IEEEyesnumber \label{x1}
\end{IEEEeqnarray*}

Substituting \eqref{x1} into \eqref{k2step2} and performing some simplification yields
\begin{IEEEeqnarray*}{rcl}
    K_2&\approx&\frac{r \omega  \Lambda ^{\alpha  \mu }  }{\Gamma (\mu +1) \Gamma \left(\mu _e\right)}\left(\frac{\Lambda _e}{\Theta }\right)^{\alpha _e\mu _e} \left(\lambda ^r \mu _r\right)^{\alpha  \mu +\alpha _e \mu _e}\\
    &\times&H_{1,1}^{1,1}\!\!\left[\frac{\Theta  \lambda ^{-r}}{\Lambda _e \mu _r}\middle|\!\!\!\begin{array}{c} \left(1,\frac{1}{\alpha _e}\right) \\ \left(r \left(\alpha  \mu +\alpha _e \mu _e\right),r\right) \\\end{array}\!\!\!\right]-\frac{r \omega   }{\Gamma \left(\mu _e\right)}\\
    &\times&\left(\frac{\Lambda _e \lambda ^r \mu _r}{\Theta }\right)^{\alpha _e \mu _e}H_{1,1}^{1,1}\!\!\left[\frac{\Theta \lambda ^{-r}}{\Lambda _e \mu _r}\middle|\!\!\!\begin{array}{c} \left(1,\frac{1}{\alpha _e}\right) \\ \left(r \alpha _e \mu _e,r\right) \\\end{array}\!\!\!\right].\IEEEyesnumber\label{k2final1}
\end{IEEEeqnarray*}

Following a similar approach to that used in scenario $\gamma_1 \to \infty$,  we can obtain the asymptotic $K_1$ and $K_2$ for the scenario when $\gamma_e \to \infty$ (the detailed derivation is omitted for brevity) as
\begin{IEEEeqnarray*}{rcl}
    K_1&\approx&-\frac{r (1-\omega ) \alpha _e }{c \Gamma (a) \Gamma (\mu ) \Gamma \left(\mu _e\right)}\left(\frac{b^r \Lambda _e \mu _r}{\Theta }\right)^{\alpha_e \mu _e}\\
    &\times&H_{3,3}^{2,2}\!\!\left[\frac{b^{-r}}{\Lambda  \mu _r}\middle|\!\!\!\begin{array}{c} (1,1),(1-\mu ,\frac{1}{\alpha }),(1+\frac{r \alpha _e \mu _e}{c},\frac{r}{c}) \\ \left(\frac{r \alpha _e \mu _e}{c},\frac{r}{c}\right),(a+\frac{r \alpha _e \mu _e}{c},\frac{r}{c}),(0,1) \\\end{array}\!\!\!\right]\IEEEyesnumber\label{pdfam}
\end{IEEEeqnarray*}
and
\begin{IEEEeqnarray*}{rcl}
    K_2&\approx&-\frac{r \omega  \alpha _e }{\Gamma (\mu ) \Gamma \left(\mu _e\right)}\left(\frac{\Lambda _e \lambda ^r \mu _r}{\Theta }\right)^{\alpha _e\mu _e}\\
    &\times&H_{2,2}^{1,2}\!\!\left[\frac{\lambda ^{-r}}{\Lambda  \mu _r}\middle|\!\!\!\begin{array}{c} (1,1),(1-\mu ,\frac{1}{\alpha }) \\ \left(r \alpha _e \mu _e,r\right),(0,1) \\\end{array}\!\!\!\right].\IEEEyesnumber
\end{IEEEeqnarray*}

It is noted that, as $\gamma_1\to \infty$, the first term in both \eqref{k1final1} and \eqref{k2final1} tends to zero, while the second term tends to a constant related to the quality of both the UWOC and the eavesdropping RF channels, which means that the secrecy outage capacity will be saturated at high transmit power $\gamma_1$. In the next section, the simulation results will again confirm this theorem. Also, the expression consisting of the second terms of \eqref{k1final1} and \eqref{k2final1} are also drawn together in the simulation, using saturation results as the legend.

\section{Numerical Results And Discussion}

In this section, we use Monte-Carlo simulations to verify the correctness of exact closed-form expressions and asymptotic expressions. Furthermore, by varying the parameter values of the $\alpha-\mu$  and the EGG models, we thoroughly investigate the relationship between the secrecy performance of the mixed RF/UWOC system and propagation medium non-linearity and the number of multipath clusters in the RF channel, and the temperature gradient and air bubbles in the UWOC channel. For simplicity, we use$[\cdot, \cdot]$ to represent the values of $[\text{air bubbles level}, \text{temperature gradient}]$ in this section.

\begin{figure}[!t]
    \includegraphics[width=.45\textwidth]{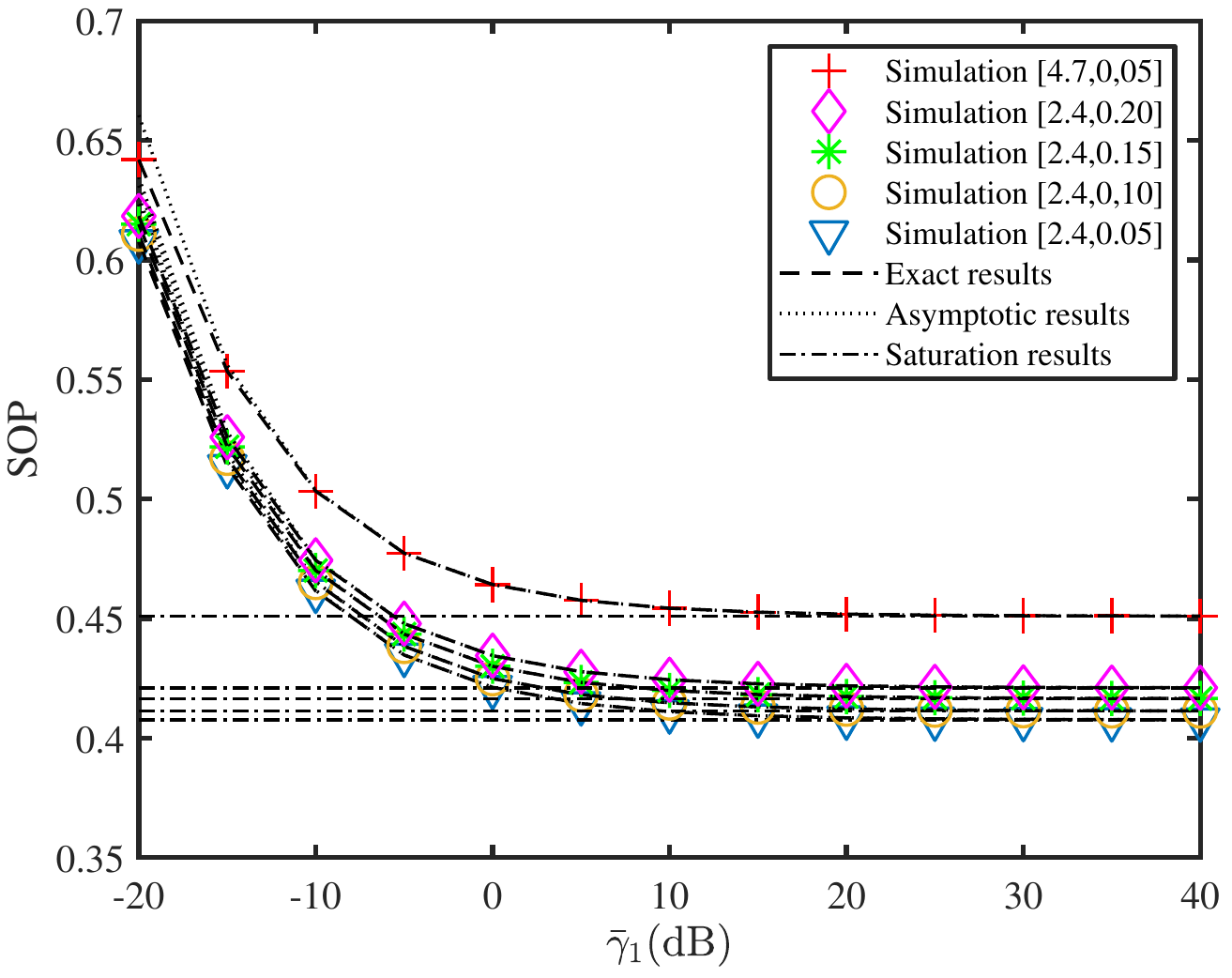}
    \centering
    \caption{SOP versus $\bar{\gamma}_1$ with various UWOC parameters and $\alpha=\alpha_e=1.2$, $\mu=\mu_e=0.5$, $R_s$=0.5, and $\bar{\gamma}_e=\bar{\gamma}_2=$ -20 dB.}
    \label{fig: 1}
\end{figure}
\begin{figure}[!t]
    \includegraphics[width=.45\textwidth]{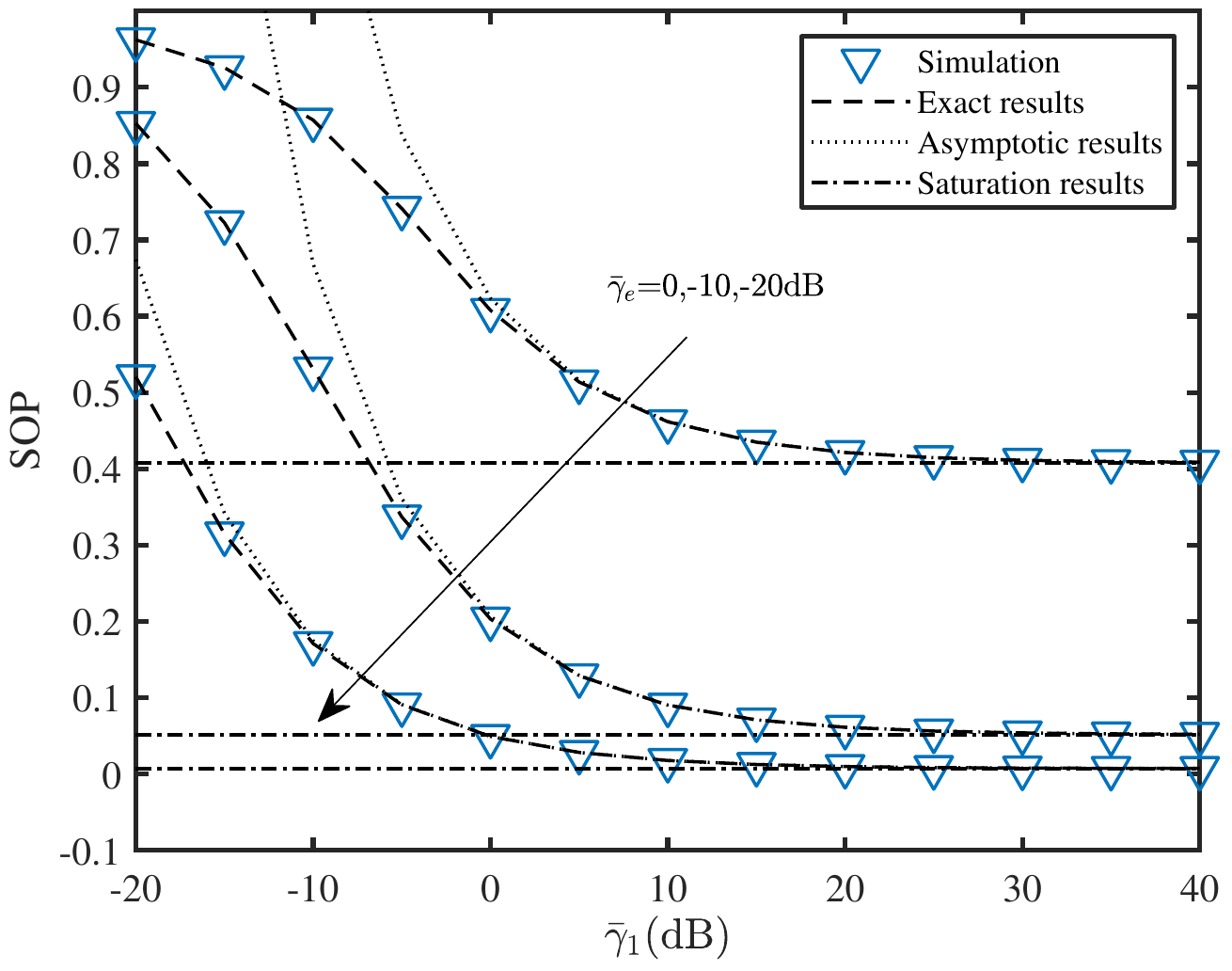}
    \centering
    \caption{SOP versus $\bar{\gamma}_1$ with various $\bar{\gamma}_e$ and UWOC parameters [2.4, 0.05], $\alpha=\alpha_e=1.2$, $\mu=\mu_e=0.5$, and $\bar{\gamma}_2$= -20 dB.}
    \label{fig: 2}
\end{figure}
\begin{figure}[!t]
    \includegraphics[width=.45\textwidth]{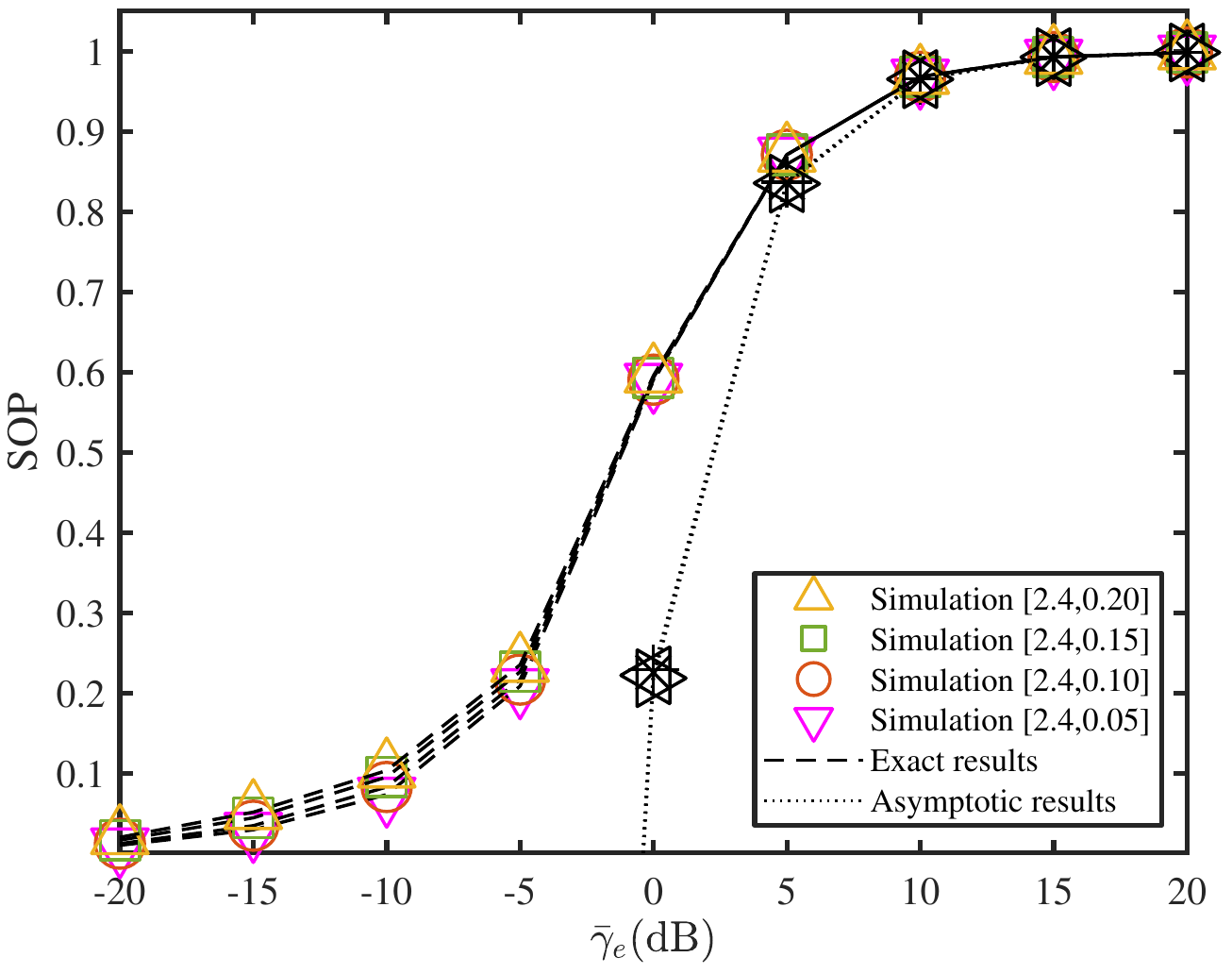}
    \centering
    \caption{SOP versus $\bar{\gamma}_e$ with various UWOC parameters and $\alpha=\alpha_e=0.9$, $\mu=\mu_e=1.5$, $R_s$=0.5, $\bar{\gamma}_1=$ 30 dB, and $\bar{\gamma}_2=$ 0 dB.}
    \label{fig: 3}
\end{figure}
\begin{figure}[!t]
    \includegraphics[width=.45\textwidth]{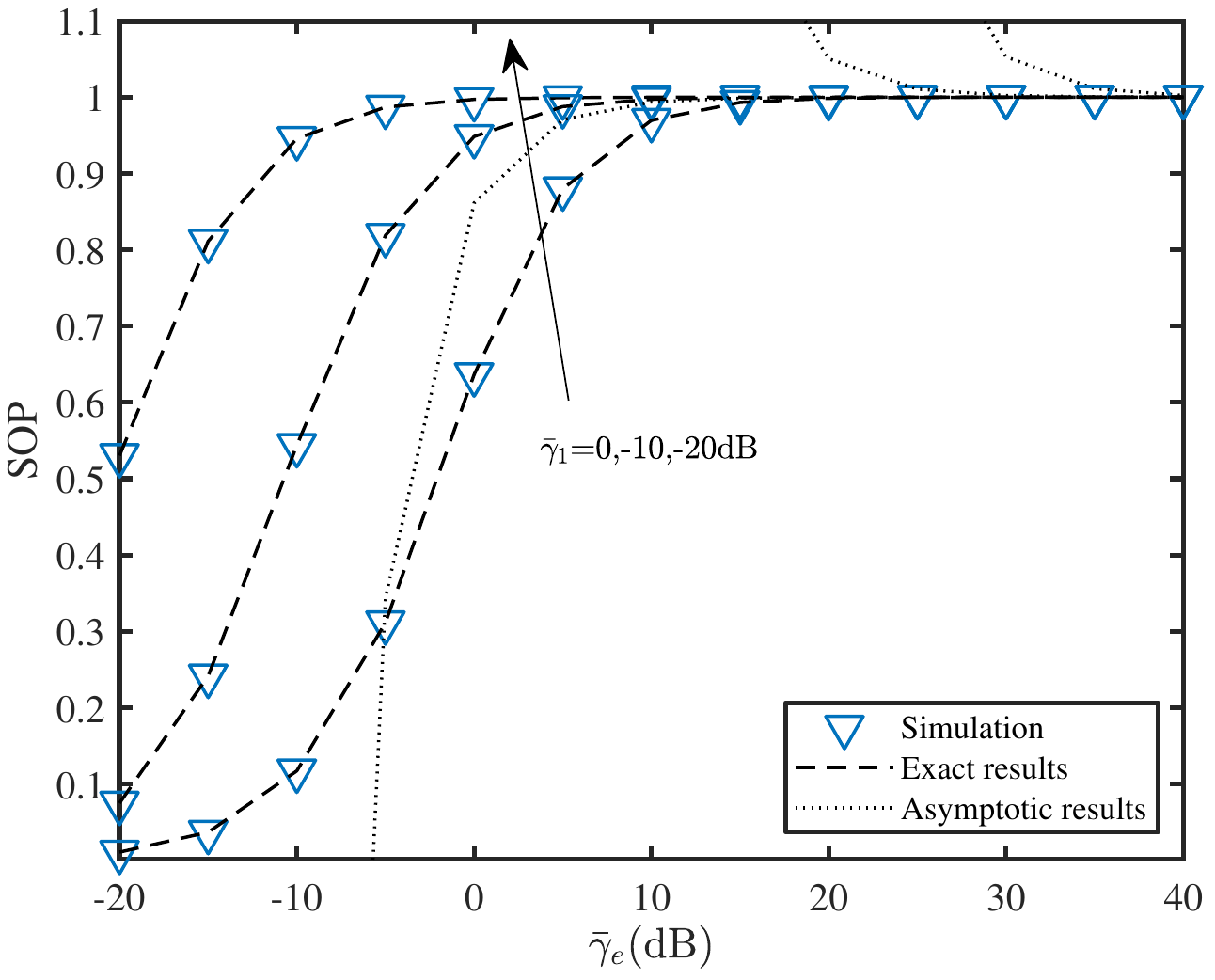}
    \centering
    \caption{SOP versus $\bar{\gamma}_e$ with various $\bar{\gamma}_1$ and UWOC parameters [2.4, 0.05], $\alpha=\alpha_e=0.9$, $\mu=\mu_e=1.5$, and $\bar{\gamma}_2$= 0 dB.}
    \label{fig: 4}
\end{figure}

Fig. 1 veriﬁes the exact and asymptotic expressions of  SOP against the SNR of SR link $\gamma_1$ over the two-hop mixed RF/UWOC system with various UWOC parameters. The average SNR of the UWOC channel is ﬁxed to $\bar{\gamma}_2=-20$ dB. The parameters of the UWOC channels for scenarios 1 to 4 are [2.4, 0.05], [2.4 0.10], [2.4, 0.15], [2.4, 0.20], and [4.7, 0.05], respectively. The SE and SR channels have the same parameters, i.e., $\alpha=\alpha_e=1.2$ and $\mu=\mu_e=0.5$. As shown in the figure, analytical and simulation results well match to each other. Moreover, when the SNR is between -20 dB and 10 dB, the SOP decreases as the SNR increases. However, from 10 dB onwards, the SOP is saturated, which conﬁrms the claims of  the theorem in the last paragraph of Section \RNum{3}. Then, from the point of view of energy efficiency, one should use the transmission power corresponding to the saturation starting point. For example, with a UWOC parameter of [4.7, 0.05], the corresponding optimal transmission power is 10 dB for the value of SOP equals to 0.45. Further, when the quality of the UWOC channel is better, the SOP is smaller. Actually, increasing the quality of the UWOC channel while keeping the quality of the eavesdropping link unchanged increases the overall capacity of the two-hop system, thereby increasing the SOP.

Fig. 2 uses the same parameters as in Fig. 1, except that only the parameters of the UWOC channel in scenario 1 are used, and the average SNR of the eavesdropping channel $\gamma_e$ is -20 dB, -10 dB, and 0 dB, respectively. As shown in the figure, when the quality of the SE channel is better, the saturation value of the SOP is larger and vice versa. In addition, the asymptotic results are very accurate from 0 dB, while the saturation results give a correct indication of the saturation value for each scenario.

Fig. 3 veriﬁes the exact and asymptotic expressions of  SOP against the SNR of SE link $\bar{\gamma}_e$ with fixed $\bar{\gamma}_1=30$ dB and various UWOC parameters. The other parameters are the same as in Fig. 1. The same principles that explain the curves in Fig. 1 also apply to explaining the curves in Fig. 3.

Fig. 4 uses the same parameters as in Fig. 2, except that $\bar{\gamma}_2$= 0 dB and $\gamma_e$ is -20 dB, -10 dB, and 0 dB, respectively. From the figure, we can observe that when the SE link quality is fixed, the better the SR link quality, the smaller the value of SOP at -20 dB and the larger the corresponding SNR value as SOP increases to 1.

\section{Conclusion}
We considered the secrecy performance of a mixed RF/UWOC system, where the EGG distribution is used for modeling the UWOC channel and $\alpha-\mu$ distribution is used for model RF links for legitimate and eavesdropping users. We derived the exact closed-form and asymptotic expressions of the secrecy outage probability and investigated the effect of channel quality on the SOP performance.


\bibliographystyle{IEEEtran}
\bibliography{IEEEabrv,ref.bib}
\end{document}